\documentclass[aps,showpacs,preprint,epsf,epsfig, nofootinbib]{revtex4-1}
\usepackage{mathrsfs}
\usepackage{dcolumn}
\usepackage{bm}
\usepackage{hyperref}
\usepackage{epsfig,graphicx,color,appendix}
\usepackage{multirow}
\usepackage{tabularx}
\usepackage{capt-of}
\usepackage{caption}
\usepackage[countmax]{subfloat}

\usepackage{threeparttable}
\usepackage{amssymb}
\usepackage{array}
\usepackage{amsmath}
\renewcommand{\thefootnote}{\fnsymbol{footnote}}
\begin{document}
\def\ben{\begin{eqnarray}}
\def\en{\end{eqnarray}}
\def\non{\nonumber}
\def\la{\langle}
\def\ra{\rangle}
\def\t{\times}
\def\pp{{\prime\prime}}
\def\nc{N_c^{\rm eff}}
\def\vp{\varepsilon}
\def\hep{\hat{\varepsilon}}
\def\a{{\cal A}}
\def\B{{\cal B}}
\def\c{{\cal C}}
\def\d{{\cal D}}
\def\e{{\cal E}}
\def\p{{\cal P}}
\def\tt{{\cal T}}
\def\up{\uparrow}
\def\dw{\downarrow}
\def\vma{{_{V-A}}}
\def\vpa{{_{V+A}}}
\def\smp{{_{S-P}}}
\def\spp{{_{S+P}}}
\def\J{{J/\psi}}
\def\ov{\overline}
\def\Lqcd{{\Lambda_{\rm QCD}}}
\def\pr{{Phys. Rev.}~}
\def\prl{{ Phys. Rev. Lett.}~}
\def\pl{{ Phys. Lett.}~}
\def\np{{ Nucl. Phys.}~}
\def\zp{{ Z. Phys.}~}
\long\def\symbolfootnote[#1]#2{\begingroup%
\def\thefootnote{\fnsymbol{footnote}}\footnote[#1]{#2}\endgroup}
\def\lsim{ {\ \lower-1.2pt\vbox{\hbox{\rlap{$<$}\lower5pt\vbox{\hbox{$\sim$}
}}}\ } }
\def\gsim{ {\ \lower-1.2pt\vbox{\hbox{\rlap{$>$}\lower5pt\vbox{\hbox{$\sim$}
}}}\ } }

\font\el=cmbx10 scaled \magstep2{\obeylines\hfill \today}

\vskip 1.5 cm

\centerline{\large\bf Magnetic Moments of Bottom Baryons : }
 \centerline{\large\bf Effective Mass and Screened Charge}
 
\small
\vskip 1.0 cm

\centerline{\bf Rohit Dhir $^1$ \symbolfootnote[2]{dhir.rohit@gmail.com}, C. S. Kim $^1$ \symbolfootnote[3]{cskim@yonsei.ac.kr} and R.C. Verma $^2$ \symbolfootnote[4] {rcverma@gmail.com} }
\medskip
\centerline{\it $^1$Department of Physics and IPAP, Yonsei University, Seoul 120-749, Korea;}
\centerline{\it $^2$Department of Physics, Punjabi University, Patiala 147-002, India.}
\bigskip
\bigskip
\begin{center}
{\large \bf Abstract}
\end{center}
We calculate the magnetic moments of low lying heavy flavor bottom baryons  using effective quark mass and shielded quark charge scheme. We obtain the magnetic moments of both $J^P= \frac{1}{2}^+$ and $J^P=\frac{3}{2}^+$ baryon states. We compare our predictions with other theoretical approaches.

\medskip
\medskip

\vfill

\newpage
 \section{Introduction} 
The study of properties of heavy flavor baryons provides
valuable insight into the nonperturbative aspects of QCD. Particularly, investigation of baryons containing bottom (\textit{b}) quark is considered to be a necessary ingredient for understanding \textit{b}-hadron phenomenology. In recent years investigations of the heavy baryon properties have become a subject of growing interest due to the experimental observation of many heavy flavor baryons \cite{1}. All spatial-ground-state baryons carrying single charm quark have already been observed, and their masses have also been measured. Many spin-$\frac{1}{2}$ \textit{b}-baryons  $\Lambda_b$, $\Sigma_b$, $\Xi_b$ and $\Omega_b$ and spin-$\frac{3}{2}$ baryon $\Sigma_b^\ast$ have also been discovered \cite{1,2,3,4,5}.  Recently, SELEX Collaboration announced the doubly heavy spin-$\frac{1}{2}$ baryon state $\Xi_{cc}^+$ with two charm quarks \cite{3}. Very recently, CMS Collaboration at CERN  observed the spin-$\frac{3}{2}$ heavy $\Xi_b^\ast$ baryon state and reported a new measurement of lifetime of $\Lambda_b$ baryon \textit{i.e.} $\tau _ {\Lambda_{b}}= 1.503 \pm 0.052$(stat.)$\pm 0.031$(syst.) ps \cite{6,7}. The hope for detection of doubly heavy and triply heavy baryons predicted by the quark model at the LHCb detector has been further raised by remarkable improvements in instrumentation and technology. 
Masses and magnetic moments serve as a rich source of information on the internal structure of hadrons. Experimentally, there exist measurements of the baryon magnetic moments of all the octet $J^P= \frac{1}{2}^+$ baryons (except for the $\Sigma^0$) and two of the magnetic moments of $J^P= \frac{3}{2}^+$ baryon decuplet \cite{2,8}.  Theoretically, there exist serious discrepancies between the quark model predictions and experimental results. Magnetic moments of heavy baryons have been considered in several theoretical approaches. Extensive literature based on  naive quark models, non-relativistic quark model (NRQM), logrithimic potential approach,  bound state approach, relativistic quark model, effective mass scheme, power-law potential model, the skyrmion model, chiral quark model, chiral perturbation theory, QCD spectral sum rules etc. \cite{9,10,11,12,13,14,15,16,17,18,19,20,21,22,23,24,25,26,27,28,29} have been employed to analyse masses and magnetic moments of heavy baryons. Due to increasing experimental activity in bottom quark sector the theoretical focus has now been shifted on the \textit{b}-baryon properties.  Most recent theoretical analyses  employ NRQM using AL1 potential \cite{30,31}, light cone QCD sum rules \cite{32,33,34}, relativistic three quark model \cite{35,36,37}, hypercentral model \cite{38} and MIT bag model \cite{39,40} to calculate the magnetic moments and radiative decays of \textit{b}-baryons. Earlier, Kumar, Dhir and Verma \cite{27} used effective quark mass and screened quark charge formalism to predict the magnetic moments of spin-$\frac{1}{2}$ charm baryons, which was later extended to spin-$\frac{3}{2}$ charmed ($C=1,2$ and $3$) baryons. In the present work, we further extend our analysis to bottom sector to determine the magnetic moments of baryons containing one or more \textit{b}-quarks in effective quark mass and screened quark charge scheme. We compare our predictions with results from other theoretical approaches.

 \section {Effective quark mass scheme}

We calculate the effective mass of the quark resulting from its interaction with the spectator quarks by single gluon exchange. Magnetic moment of baryons are obtained by using effective quark masses. The baryon mass is taken to be the sum of the quark masses plus spin-dependent hyperfine interaction \cite{16},

\ben M_{B} =\sum _{i}m_{i}^{\mathscr{E}}  = \sum _{i}m_{i}  +\sum _{i<j}b_{ij}     \mathbf{s}_{i}  \cdot \mathbf{s}_{j} ,\en 
where, $\mathbf{s}_i$ and $\mathbf{s}_j$ are the spin operators of the \textit{i}th and \textit{j}th quark, respectively; $m_{i}^{\mathscr{E}} $ denote the effective mass of the quark inside a baryon and $b_{ij} $ is given by 

\ben b_{ij} = \frac{16\pi   \alpha _{s} }{9m_{i} m_{j} } \left\langle\Psi _{0} \left|\delta ^{3} (\vec{r})\right|\Psi _{0}\right\rangle \en     for baryons $B(qqq)$ where $\Psi_0$ is the baryon wave function.            

  There may also be a spin independent interaction term, the effect of which can be approximated by the renormalization of quark masses. Thus, the mass of the quark inside the baryon $B (123)$ may get modified due to its interaction with other quarks. For quarks 1 and 2 to be identical, we write 

\begin{equation}    
m_{1}^{\mathscr{E}}=m_{2}^{\mathscr{E}} =m+\alpha   b_{12} +\beta   b_{13}   , 
\end{equation} 

\begin{equation}   
m_{3}^{\mathscr{E}} =m_{3} +2\beta   b_{13} ,                                                  
\end{equation} 
where we use $m_{1} =m_{2} =m$ and $b_{13} =b_{23} $; $\alpha $ and $\beta $ are the parameters to be determined as follows.                                          

 For $J^P=\frac{1}{2}^+$ states,  
\ben M_{B} =\sum _{i}m_{i}  +\sum _{i<j}b_{ij}     \mathbf{s}_{i}  \cdot \mathbf{s}_{j} ,\en
simplified to,
\begin{equation}    
M_{B_{\frac{1}{2}^{+} } } =2m+m_{3} +\frac{b_{12} }{4} -b_{13}   ,
\end{equation} 
for
\begin{equation}    
\mathbf{s}_{1} \cdot \mathbf{s}_{2} =\frac{1}{4}   ,  \mathbf{s}_{1} \cdot \mathbf{s}_{3} =\mathbf{s}_{2} \cdot \mathbf{s}_{3} =-\frac{1}{2}   . 
\end{equation} 
thereby giving, \begin{equation}    
\alpha =\frac{1}{8}   ~{\rm and} ~   \beta =-\frac{1}{4}   , 
\end{equation} 
Equation (1) can be written in generalized form for $J^P=\frac{1}{2}^+$ baryons as 
\begin{equation}    
M_{B_{\frac{1}{2}^{+} } } =m_{1} +m_{2} +m_{3} +\frac{b_{12} }{4} -\frac{b_{23} }{2} -\frac{b_{13} }{2}   ,   
\end{equation} 
where 1, 2, 3 represents \textit{u},\textit{ d}, \textit{s},\textit{ c}, and \textit{b} quarks. Following the formalism described above, for $J^P=\frac{3}{2}^+$ baryons we get,
\begin{equation}    
 M_{B_{\frac{3}{2}^{+} } } =m_{1} +m_{2} +m_{3} +\frac{b_{12} }{4} +\frac{b_{23} }{4} +\frac{b_{13} }{4},  
\end{equation} 
for \begin{equation}    
\alpha =  \beta =\frac{1}{8}.
\end{equation} 
The parameterization used here seems to go beyond the leading order in quark mass splitting because the $m_i$ term appears as $1/m_i m_j$ through the hyperfine interaction. However, higher order effects are at least partially absorbed in the nonlinear fitting of the $m_i$. It has been shown \cite{14,15} that contributions from new nonlinear terms must be small because the fitted masses satisfy the  Gell-Mann--Okubo mass formula, which is exact to leading order in the quark mass splitting. Therefore, the effective quark mass defined here is equivalent to first order in baryon mass splitting to the leading order parameterization of the baryon masses in chiral perturbation theory \cite{14}. 

Values of quark masses and hyperfine interaction terms $b_{ij} $ are obtained from the known iso-multiplet masses. We wish to point out that as compared to our previous work the $b_{ij}$'s are obtained here in a more realistic manner corresponding to strange, charm and bottom mass scales. \textit{N}, $\Delta $, $\Lambda $, $\Lambda _{c} $ and $\Lambda _{b} $  gives,

\begin{equation}    
\begin{array}{c} {m_{u}   =  m_{d}   =  362    ~{\rm MeV},  m_{s}   =  539~{\rm MeV},    m_{c}   =  1710 ~ {\rm MeV},} \\ {   m_{b}   =  5043 ~ {\rm MeV~and~}   b_{uu}^{}   =  b_{ud}^{}   =  b_{dd}^{} =  196 ~  {\rm MeV}.} \end{array} 
\end{equation} 
From $\Sigma $ and $ \Omega $ we obtain:
\ben    b_{us}  & = & b_{ds}   =118  ~{\rm MeV}, \\ \non
    b_{ss}   &=&   76 ~ {\rm MeV}. \en
In charm sector, $\Sigma_c$ gives
\ben b_{uc}   =  b_{dc}   =    28 ~ {\rm MeV},\en
which in turn yields,
 \ben b_{sc}  &=&\left(  \frac{m_{_{u} }  }{m_{s}   } \right)b_{uc}  =  19 ~ {\rm MeV},\\ \non 
  b_{cc}  & =&  \left(  \frac{m_{u} }{m_{c} } \right) b_{uc}   =  6 ~ {\rm MeV}.\en 
In bottom sector, $\Sigma_b$ and $\Sigma^{*}_b$, leads to
\ben    b_{ub} &= & b_{db} =7~{\rm MeV},\en 
which gives, \ben b_{sb} &=&\left( \frac{m_{u} }{m_{s}   } \right) b_{ub} = 5~{\rm MeV},\\ \non
  b_{cb} &=&\left( \frac{m_{u} }{m_{c}    } \right) b_{ub} = 1.5~{\rm MeV},\\ \non
  b_{bb} &=&\left(  \frac{m_{u} }{m_{b} } \right) b_{ub} =0.5~{\rm MeV}.  
\en 
Assuming matrix element of spatial part of baryonic wave function to be flavor and spin-independent, we have now extracted ratios of $\alpha_s$ for different quark mass scales \textit{i.e.}  
\ben \frac{\alpha_s(ss)}{\alpha_s(uu)} &= 0.86,~ \frac{\alpha_s(cc)}{\alpha_s(ss)}=0.80 ,~\frac{\alpha_s(bb)}{\alpha_s(cc)}=& 0.75, \\ \non
\frac{\alpha_s(us)}{\alpha_s(uu)}&= 0.90,~\frac{\alpha_s(uc)}{\alpha_s(uu)}= 0.68,~\frac{\alpha_s(ub)}{\alpha_s(uu)}=&0.51 .\en 
However, spatial part of the hadron wave function may show flavour dependence, since size of the hadron may vary with quark flavors. In meson sector, leptonic decay width show flavor dependence of spatial part of the wave function. Similar $\alpha_s$ ratios at other mass scales can be obtained from $b_{ij}$ relations.

Using these values of quark masses and hyperfine interaction terms  $b_{ij}$, we can obtain the effective quark masses for $J^{P}   =  \frac{1}{2}^{+} $ and $J^{P}   =  \frac{3}{2}^{+} $ baryons. 

\section{ Magnetic Moments of  $\pmb{(J^{P}   =  \frac{1}{2}^{+} )}$ Baryons in effective Mass Scheme}

 In the present scheme, magnetic moments of $J^{P}   =  \frac{1}{2}^{+} $ baryons are obtained by sandwiching the following magnetic moment operator between the appropriate baryon wave functions: 

\begin{equation}  
\pmb{\mu }=\sum _{i}\mu _{i}^{\mathscr{E}}  \pmb{\sigma }_{i} ,                                                 
\end{equation} 
where 
 \ben \mu _{i}^{\mathscr{E}} =\frac{e_{i} }{2m_{i}^{\mathscr{E}} } ,\en    
 for \textit{i = u, d, s,c }and\textit{ b}; $e_i$ represent the quark charge. Expressions for magnetic moments of $J^P=\frac{1}{2}^+$ bottom baryons are given in Table\ref{t1}. We also obtain magnetic transition moments $B_{1/2^+}^{\prime} (123) \to B_{1/2^+}(123)$ following the general expression given by 
\ben \mu^{\mathscr{E}}_{B^{\prime}_{1/2} \to B_{1/2}} &=& \mu^{\mathscr{E}}_{B^{\prime}_{1/2}}(123) - \mu^{\mathscr{E}}_{B_{1/2}}(123) \non \\ 
&=& \left[\mu^{\mathscr{E}}(2)-\mu^{\mathscr{E}}(1)\right]/\sqrt{3}. \en

\begin{table}
\centering

\captionof{table} {Expressions for magnetic moments of $(J^P = \frac{1}{2}^+)$ baryons using effective quark masses (in nuclear magneton).}
\label{t1}
  \begin{tabular} {|c |c |} \hline 
\textbf{Particles} & \textbf{Magnetic Moment} \\ \hline 
\multicolumn{2}{| l|}{Singly heavy} \\ \hline 
$\Sigma _{b}^{+} $ & $(-\mu _{b} +4\mu _{u} )/3$ \\ \hline 
$\Sigma _{b}^{0} $ & $(-\mu _{b} +2\mu _{d} +2\mu _{u} )/3$ \\ \hline 
$\Sigma _{b}^{-} $ & $(-\mu _{b} +4\mu _{d} )/3$ \\ \hline 
$\Xi _{b}^{'0} $ & $(-\mu _{b} +2\mu _{s} +2\mu _{u} )/3$ \\ \hline 
$\Xi _{b}^{'-} $ & $(-\mu _{b} +2\mu _{d} +2\mu _{s} )/3$ \\ \hline
$\Lambda _{b}^{0} $ & $\mu _{b} $ \\ \hline 
$\Xi ^{0}_{b} $ & $\mu _{b} $ \\ \hline 
$\Xi ^{-} _{b} $ & $\mu _{b} $ \\ \hline 
$\Omega _{b}^{-} $ & $(-\mu _{b} +4\mu _{s} )/3$ \\ \hline 
\multicolumn{2}{|l|}{Doubly heavy} \\ \hline 

$\Xi ^{'+} _{cb} $ & $\mu _{u} $ \\ \hline 
$\Xi ^{'0} _{cb} $ & $\mu _{d} $ \\ \hline 
$\Omega^{'0}_{cb}  $ & $\mu _{s} $ \\ \hline 
$\Xi _{cb}^{+} $ & $(2\mu _{b} +2\mu _{c} -\mu _{u} )/3$ \\ \hline 
$\Xi _{cb}^{0} $ & $(2\mu _{b} +2\mu _{c} -\mu _{d} )/3$ \\ \hline 
$\Omega _{cb}^{0} $ & $(2\mu _{b} +2\mu _{c} -\mu _{s} )/3$ \\ \hline 
$\Xi _{bb}^{0} $ & $(4\mu _{b} -\mu _{u} )/3$ \\ \hline 
$\Xi _{bb}^{-} $ & $(4\mu _{b} -\mu _{d} )/3$ \\ \hline 
$\Omega _{bb}^{-} $ & $(4\mu _{b} -\mu _{s} )/3$ \\ \hline  
\multicolumn{2}{|l |}{Triply Heavy} \\ \hline 

$\Omega _{ccb}^{+} $ & $(-\mu _{b} +4\mu _{c} )/3$ \\ \hline
$\Omega _{cbb}^{0} $ & $(4\mu _{b} -\mu _{c} )/3$ \\ \hline 
\end{tabular}
\end{table}

We wish to point out that we use simplistic approach based on the non-relativistic magnetic moment. It has been shown by Morpurgo \cite{14} that for the static properties of baryons the non-relativistic constituent quark model approach is completely equivalent to a parametrization of relativistic field theory of strong interactions in a spin-flavor basis. A similar argument connecting constituent quark model and HBChPT is floated by Durand \textit{et al.} \cite{15}. Remarkable success of the usual additive quark model is thus a consequence of the relative smallness of the non-additive two- and three-body operators arising from spin-dependent interactions. So, the use of the one-body operator is justified in light of the decoupling of spatial and spins of the ground state baryon wave functions \cite{14,15}. 
 
   We use (9) to obtain more general expressions for effective masses of the quarks inside the baryon as follows:

\begin{enumerate}
\item  For (\textit{aab})-type baryons with quarks 1 and 2 being identical,
\begin{equation}    
m_{1}^{\mathscr{E}} =m_{2}^{\mathscr{E}} =m+\frac{b_{12} }{8} -\frac{b_{13} }{4} ,   
\end{equation} 
and
       \begin{equation}  m_{3}^{\mathscr{E}} =m_{3} -\frac{b_{13} }{2}   ~{\rm for}~ 1=2 \ne 3. 
       \end{equation}

\item  The baryonic states with three quarks of different flavor (\textit{abc}) can have antisymmetric $\Lambda_{\left[12\right]3}$-type  and symmetric $\Sigma_{\{12\}3}$-type flavor configurations under the exchange of quarks 1 and 2. 
\begin{enumerate}
    \item For (\textit{abc}) $\Lambda $-type baryons,
\ben    
m_{1}^{\mathscr{E}} =m_1 -\frac{3b_{12} }{8} , \\  
m_{2}^{\mathscr{E}} =m_2-\frac{3b_{21} }{8} , 
\en 
and 
\begin{equation}
 m_{3}^{\mathscr{E}} =m_{3}     ~{\rm for}~  1\ne 2\ne 3. 
\end{equation}

\item  For (\textit{abc}) $\Sigma $\textit{-}type baryons,
\begin{equation}    
m_{1}^{\mathscr{E}} =m_{1} +\frac{b_{12} }{8} -\frac{b_{13} }{4} ,   
\end{equation} 
\begin{equation}   m_{2}^{\mathscr{E}} =m_{2} +\frac{b_{12} }{8} -\frac{b_{23} }{4} ,   
\end{equation} 
and
\begin{equation}    
m_{3}^{\mathscr{E}} =m_{3} -\frac{b_{23} }{4} -\frac{b_{13} }{4}       ~{\rm for}~  1\ne 2\ne 3. 
\end{equation} 
  \end{enumerate}
\end{enumerate}

Using these relations we obtain the effective quark masses for $J^{P}   =  \frac{1}{2}^{+} $ baryons as follows:

\begin{enumerate}
\item  For singly heavy baryons, 
\[m_{u}^{\Lambda _{b} } =m_{d}^{\Lambda _{b} } = 288 {~\rm MeV}, m_{b}^{\Lambda _{b} }=  5043 ~\rm MeV;\]

                                    \[m_{u}^{\Xi_{b} } =m_{d}^{\Xi_{b} } = 318 {~\rm MeV}, m_{s}^{\Xi_{b} }  =495{~\rm MeV}, m_{b}^{\Xi_{b} } =5043 {~\rm MeV};\]                      
                                     
                                     \[m_{s}^{\Omega _{b} } =547 {~\rm MeV},m_{b}^{\Omega _{b} } =5041{~\rm MeV};\]
                                     
                                     \[m_{u}^{\Sigma _{b} } =m_{d}^{\Sigma _{b} } = 385 {~\rm MeV}, m_{b}^{\Sigma _{b} } =5039{~\rm MeV};\]
                                      \ben m_{u}^{\Xi^{'} _{b} } =m_{u}^{\Xi^{'} _{b} }= 375 {~\rm MeV}, m_{s}^{\Xi^{'} _{b} } =553{~\rm MeV}, m_{b}^{\Xi^{'} _{b} } =5040{~\rm MeV};\en
\item  For doubly heavy baryons,
\[ m_{u}^{\Xi_{cb}  } =m_{d}^{\Xi_{cb}  } = 353 {~\rm MeV}, m_{c}^{\Xi_{cb} } =1703{~\rm MeV},m_{b}^{\Xi_{cb} } =5041{~\rm MeV}.\]

\[ m_{s}^{\Omega_{cb} } =533{~\rm MeV}, m_{c}^{\Omega_{cb} } =1705{~\rm MeV}, m_{b}^{\Omega_{cb} } =5042 {~\rm MeV};\]

                                 \[ m_{u}^{\Xi^{'} _{cb}  } =m_{u}^{\Xi^{'} _{cb}  } = 362 {~\rm MeV}, m_{c}^{\Xi^{'} _{cb}  } =1709{~\rm MeV}, m_{b}^{\Xi^{'} _{cb}  } =50421{~\rm MeV}; \]
                                                    
 \[m_{s}^{\Omega _{cb}^{'} } =539 {~\rm MeV}, m_{c}^{\Omega _{cb}^{'} } =1709{~\rm MeV}, m_{b}^{\Omega _{cb}^{'} } =5042{~\rm MeV};\]
 
 \[m_{u}^{\Xi _{_{bb} } } =m_{d}^{\Xi _{_{bb} } } =358 {~\rm MeV}, m_{b}^{\Xi _{_{bb} } } =5041{~\rm MeV};\]
 \ben m_{s}^{\Omega _{bb} } =537 {~\rm MeV}, m_{b}^{\Omega _{bb} } = 5042{~\rm MeV};\en
\item  For triply heavy baryons,
\[ m_{c}^{\Omega _{ccb} } =1710{~\rm MeV}, m_{b}^{\Omega _{ccb} } =5042 {~\rm MeV}.\]
\ben m_{c}^{\Omega _{cbb} } =1709{~\rm MeV}; m_{b}^{\Omega _{cbb} } =5043{~\rm MeV}.\en                                    
\end{enumerate}
\begin{table}
\centering

\captionof{table} {Masses of $(J^P = \frac{1}{2}^+)$ bottom baryons using effective quark masses (in GeV).}
\label{mt1}
  \begin{tabular} {|c |c |c|c|c|c|c|c|} \hline 
\textbf{Particles} & \textbf{Masses } &\textbf{\cite{30}} &\textbf{\cite{33, 34}}& \textbf{\cite{36}}& \textbf{\cite{38}}$^@$&\textbf{\cite{41}}&\textbf{Expt.\cite{2}}\\ \hline
\multicolumn{8}{| l|}{Singly heavy} \\ \hline 
$\Sigma _{b}  $ & 5.808  & 5.808& - & 5.82 & 5.858  & 5.805 & 5.813$^*$   \\ \hline 
$\Xi _{b}^{'} $ & 5.967 &5.946& 5.945 &5.94 &- &   5.937  &- \\ \hline
$\Lambda _{b} $ & 5.620   &5.643&-&  5.624 & 5.693 &   5.622 &5.619$^*$    \\ \hline 
$\Xi _{b} $ & 5.855  & 5.850&5.824 & 5.89  & 5.922 & 5.812& 5.790 \\ \hline 
$\Omega _{b} $ & 6.135  &6.034&-  &6.04 & 6.052 & 6.065& 6.071 $\pm$0.40 \\ \hline 
\multicolumn{8}{|l|}{Doubly heavy} \\ \hline 
$\Xi ^{'} _{cb} $ & 7.114 &6.948& 6.97 $\pm$ 0.20 & 6.85 &  -& 6.963&  \\ \hline 
$\Omega^{'}_{cb}  $ & 7.291&7.009& 6.80 $\pm$ 0.30&   6.93 & - & 7.116   & \\ \hline 
$\Xi _{cb} $ & 7.097 &6.919& 6.72$\pm$ 0.20 &  6.82 & 6.887 &6.933&  \\ \hline 
$\Omega _{cb} $ & 7.281  & 6.986& 6.75$\pm$ 0.30& 6.91& 6,952   &7.088& \\ \hline 
$\Xi _{bb} $ & 10.440 &10.197&9.96$\pm$ 0.90 & 10.10& 10.162   &10.202&  \\ \hline 
$\Omega _{bb} $ &  10.620 &10.260& 9.97$\pm$ 0.90& 10.18  & 10.220 & 10.359&  \\ \hline  
\multicolumn{8}{|l |}{Triply Heavy} \\ \hline 

$\Omega _{ccb}  $ &  8.463&- &8.50 $\pm$ 0.12 & 8.00 &8.172&& \\ \hline
$\Omega _{cbb} $ &  11.795 &-&11.72 $\pm$ 0.16 & 11.50 &11.447 && \\ \hline 
\end{tabular}
\begin{tablenotes}
      \footnotesize 
      \item * used as input
      \item @ averaged
    \end{tablenotes}
\end{table}

Using these effective quark masses, we obtain masses of baryon iso-multiplets as shown column 2 of Table \ref{mt1}. Further, we calculate the magnetic moments of $J^P=\frac{1}{2}^+$ baryons as given in column 2 of Table \ref{t2}. To determine magnetic $B_{1/2}^\prime \to B_{1/2}$ transition moments, we take the geometric average of effective quark masses of constituent quarks of initial and final state baryons \textit{i.e.} $m_i^{B' \to B}=\sqrt{m_i^{B'} m_i^{B}}$. For the sake of comparison we also give the results of the different models namely MIT bag model \cite{39}, hyper central potential \cite{38}, relativistic three quark \cite{36}, light cone QCD sum rules \cite{32}, NRQM using AL1 potential \cite{30}, power-law potential \cite{17} etc.
\begin{table}
\centering
\captionof{table} {Magnetic moments of $(J^P= \frac{1}{2}^+)$ bottom baryons using effective quark masses (in nuclear magneton).}
\label{t2}
\begin{tabular}{|c|>{\centering}p{2.6cm}|>{\centering}p{2.8cm}|c|c|c|c|c|c|} \hline 
\multirow{2}{*}{ \textbf{ Baryons}} & \multicolumn{2}{c|}{\textbf{This Work}} & \multirow{2}{*}{\textbf{ \cite{39}}} & \multirow{2}{*}{\textbf{ \cite{12,39}}} &\multirow{2}{*}{ \textbf{\cite{38}}} & \multirow{2}{*}{\textbf{ \cite{36} }} & \multirow{2}{*}{\textbf{\cite{30} }} & \multirow{2}{*}{\textbf{ \cite{17} }}  \\ \cline{2-3} 
 &  \textbf{Effective Quark Mass}  &  \textbf{Screened Quark Charge}  &  &  &  &  &  &    \\ \hline 
\multicolumn{9}{|l|}{Singly heavy} \\ \hline 
 $\Sigma _{b}^{+} $  &  2.190  &  2.177  &  $1.622$  &  $2.50$  &  2.229  &  $2.07$  &  ---  &  $2.575$  \\ \hline 
 $\Sigma _{b}^{0} $  &  0.563  &  0.533  &  $0.422$  &  $0.64$  &  0.591  &  $0.53$  &  --- &  $0.659$  \\ \hline 
 $\Sigma _{b}^{-} $  &  -1.064  &  -1.110  &  -$0.778$  &  -$1.22$  &  -1.047  &  -$1.01$  &  ---  &  -$1.256$  \\ \hline 
 $\Xi _{b}^{'0} $  &  0.756  &  0.676  &  $0.556$  &  $0.90$  &  0.766  &  $0.66$  &  ---  &  0.930  \\ \hline 
 $\Xi _{b}^{'-} $  &  -0.913  &  -0.996  &  -$0.660$  &  -$1.02$  &  -0.902  &  -$0.91$  &  ---  &  -0.985  \\ \hline 
 $\Omega _{b}^{-} $  &  -0.741  & -0.863  &  -$0.545$  &  -$0.79$  &  -0.960  &  -$0.82$  &  ---  &  -$0.714$  \\ \hline 
 $\Lambda _{b}^{0} $  &  -0.062  &  -0.060  &  -$0.066$  &  \textbf{ -$0.06$ }  &  - 0.064  &  -$0.06$  &  --- &  ---  \\ \hline 
 $\Xi _{b}^{0} $  &  -0.062  &  -0.060  &  -0.100  &  -0.110  &  ---  &  -0.06  &  ---  &  ---  \\ \hline 
 $\Xi _{b}^{-} $  &  -0.062  &  -0.066  &  -0.063  &  -0.050  &  ---  &  --0.06  &  ---  &  ---  \\ \hline 
 \multicolumn{9}{|l|}{ Doubly heavy } \\ \hline 
 $\Xi _{cb}^{'+} $  &  1.729 &  1.718  &  $1.093$  &  $1.71$  &  ---  &  $1.520$  &  1.990  &  1.525  \\ \hline 
 $\Xi _{cb}^{'0} $  &  -0.864  &  -0.817  &  -$0.236$  &  -$0.53$  &  ---  &  -$0.76$  &  -0.993  &  -0.390  \\ \hline 
 $\Omega _{cb}^{'0} $  &  -0.580  &  -0.621  &  $-0.106$  &  $-0.27$  &  ---  &  -$0.61$  & -0.542  &  -$0.119$  \\ \hline
 $\Xi _{cb}^{+} $  &  -0.387  &  -0.369  &  -$0.157$  &  -$0.25$  &  -0.400  &  -$0.12$  &  -0.475  &  ---  \\ \hline 
 $\Xi _{cb}^{0} $  &   0.499  &  0.480  &  -0.068  &  -0.13  &  0.477  &  $0.42$  &  0.518  &  ---  \\ \hline 
 $\Omega _{cb}^{0} $  &   0.399  &  0.407  &  0.034  &   0.08  &  0.397  &  $0.45$  &  0.368  &  ---  \\ \hline 
  $\Xi _{bb}^{0} $  &  -0.665  &  -0.630  &  -$0.432$  &  -$0.70$  &  -0.657  &  -$0.53$  &  -$0.742$  &  -$0.722$  \\ \hline 
 $\Xi _{bb}^{-} $  &  0.208  &  0.215  &  $0.086$  &  $0.23$  &  0.190  &  $0.18$  &  $0.251$  &  $0.236$ \\ \hline 
 $\Omega _{bb}^{-} $  &  0.111  &  0.138  &  $0.043$  &  $0.12$  &  0.109  &  $0.04$  &  $0.101$  &  $0.100$  \\ \hline
\multicolumn{9}{|l|}{ Triply heavy } \\ \hline 
  $\Omega _{ccb}^{+} $  &  0.508  &  0.522  &  $0.505$  &  $0.54$  &  ---  &  $0.53$  &  --- &  $0.476$  \\ \hline  
 $\Omega _{cbb}^{0} $  &  -0.205  &  -0.200  &  -$0.205$  &  -$0.21$  &  ---  &  -$0.20$  &  ---  &  -$0.197$  \\ \hline 
\multicolumn{9}{|l|}{ ($\frac{1}{2}^{\prime +}  \to \frac{1}{2}^{+}$) transition moments } \\ \hline 
 $\left|\Sigma _{b}^{0} \to \Lambda _{b}^{0} \right|$  &  1.627  &  1.535  &  $1.052$  &  $1.61$  &  ---  &  ---  &  ---  &  ---  \\ \hline 
 $\left|\Xi _{b}^{'0} \to \Xi _{b}^{0} \right|$  &  1.392  &  1.354  &  $0.917$  &  $1.41$  &  ---  &  ---  &  ---  &  ---  \\ \hline 
 $\left|\Xi _{b}^{'-} \to \Xi _{b}^{-} \right|$  &  0.178  & 0.142  &  $0.082$  &  $0.16$  &  ---  &  ---  &  ---  &  ---  \\ \hline 
 $\left|\Xi _{bc}^{'+} \to \Xi _{bc}^{+} \right|$  &  0.247  &  0.250  &  $0.277$  &  $0.62$  & ---  &  ---  &  ---  &  ---  \\ \hline 
 $\left|\Xi _{bc}^{'0} \to \Xi _{bc}^{0} \right|$  & 0.247  &  0.242  &  $0.508$  &  $0.70$  &  ---  &  ---  &  ---  &  ---  \\ \hline 
 $\left|\Omega _{bc}^{'0} \to \Omega _{bc}^{0} \right|$  & 0.247  &  0.243  &  $0.443$  &  $0.56$  &  ---  &  ---  &  ---  &  ---  \\ \hline 
\end{tabular}

\end{table}

It may be noted that notations of primed and unprimed states for the single heavy  $(q q Q)$ baryons $\Xi_{Q}$ and $\Xi _{Q}^{\prime }$ are used as per convention that the physical $\Xi _{Q}$ state contains a pair of light quarks $ [q_{1}q_{2}]$ mostly in a spin $S=0$ (antisymmetric) state where $q_{i}$ denotes the light and $Q$ the heavy quarks \cite{2}. The other state in which the light quark pair $ [q_{1}q_{2}]$ is mostly in spin triplet $S=1$ (symmetric) state is denoted as $\Xi _{Q}^{\prime }$. On the other hand, some complications arise in case of doubly
heavy $(q Q  Q)$ baryons $\Xi _{bc},\Xi _{bc}^{\prime }$ and $\Omega _{bc},\Omega
_{bc}^{\prime }$. These states are identified by the set of quantum numbers $(J^P,S_d)$ where $S_d$ is spin of heavy diquark. The spins of the two heavy quarks are coupled to form ($S_d=0$) antisymmetric spin configuration of diquark $[Q_{1}Q_{2}]$ and ($S_d=1$) symmetric spin configuration of diquark $\{Q_{1}Q_{2}\}$. In literature\cite{30,32,33,34,37,38,40,41}, the standard convention is to denote the symmetric heavy diquark state as unprimed $\left| B \right\rangle$ state and antisymmetric one as $\left| B^{\prime}\right\rangle $. In addition, the wave function mixing between  $\left| B \right\rangle$ and $\left| B^{\prime}\right\rangle $ states have also been considered in \cite{17,37,39}, which we have ignored in present analysis.

 \section{ Magnetic Moments of $(J^{P}   =  \frac{3}{2}^{+} )$ Baryons in effective Mass Scheme}

Proceeding in a way similar to $J^P=\frac{1}{2}^+$ , the magnetic moments of  baryons are obtained by sandwiching the magnetic moment operator (14) i.e. $ 
\pmb{\mu }=\sum _{i}\mu _{i}^{\mathscr{E}}  \pmb{\sigma }_{i}$, between the appropriate baryon wave functions, where     for \textit{i = u, d, s, c }and \textit{b.} Expressions for magnetic moments of bottom baryons are given in Table \ref{t3}. 
\begin{table}
\centering
\captionof{table} {Expressions for magnetic moments $(J^P = \frac{3}{2}^+)$ bottom baryons using effective quark masses (in nuclear magneton).}
\label{t3}
\begin{tabular}{|c |c|} \hline 
\textbf{Particles} & \textbf{ Magnetic Moment} \\ \hline 
\multicolumn{2}{|l|}{Singly heavy} \\ \hline 
$\Sigma _{b}^{*+} $ & $(\mu _{b} +2\mu _{u} )$ \\ \hline 
$\Sigma _{b}^{*0} $ & $(\mu _{b} +\mu _{d} +\mu _{u} )$ \\ \hline 
$\Sigma _{b}^{*-} $ & $(\mu _{b} +2\mu _{d} )$ \\ \hline 
$\Xi _{b}^{*0} $ & $(\mu _{b} +\mu _{s} +\mu _{u} )$ \\ \hline 
$\Xi _{b}^{*-} $ & $(\mu _{b} +\mu _{d} +\mu _{s} )$ \\ \hline
$\Omega _{b}^{*-} $ & $(\mu _{b} +2\mu _{s} )$ \\ \hline 

\multicolumn{2}{|l|}{Doubly heavy} \\ \hline
$\Xi _{cb}^{*+} $ & $(\mu _{b} +\mu _{c} +\mu _{u} )$ \\ \hline 
$\Xi _{cb}^{*0} $ & $(\mu _{b} +\mu _{c} +\mu _{d} )$ \\ \hline 
$\Omega _{cb}^{*0} $ & $(\mu _{b} +\mu _{c} +\mu _{s} )$ \\ \hline  
$\Xi _{bb}^{*0} $ & $(2\mu _{b} +\mu _{u} )$ \\ \hline 
$\Xi _{bb}^{*-} $ & $(2\mu _{b} +\mu _{d} )$ \\ \hline 
$\Omega _{bb}^{*-} $ & $(2\mu _{b} +\mu _{s} )$ \\ \hline 
\multicolumn{2}{|l|}{Triply heavy} \\ \hline
$\Omega _{ccb}^{*+} $ & $(\mu _{b} +2\mu _{c} )$ \\ \hline
$\Omega _{cbb}^{*0} $ & $(2\mu _{b} +\mu _{c} )$ \\ \hline   
$\Omega _{bbb}^{*-} $ & $(3\mu _{b} )$ \\ \hline 
\end{tabular}
\end{table}

In order to calculate the magnetic moments, we determine the effective quark masses of $(J^{P}   =  \frac{3}{2}^{+} )$ baryons from the following relations derived from (10):

\begin{enumerate}
\item  For (\textit{aab})-type baryons,
\begin{equation}    
m_{1}^{\mathscr{E}} =m_{2}^{\mathscr{E}} =m+\frac{b_{12} }{8} +\frac{b_{13} }{8} ,   
\end{equation} 
 and                 \ben 
                 m_{3}^{\mathscr{E}} =m_{3} +\frac{b_{13} }{4}   {~\rm for}~ 1=2\ne 3.\en
 \item  For (\textit{abc})\textit{-}type  baryons,
\begin{equation}    
m_{1}^{\mathscr{E}} =m_{1} +\frac{b_{12} }{8} +\frac{b_{13} }{8} ,   
\end{equation} 
\begin{equation}    
m_{2}^{\mathscr{E}} =m_{2} +\frac{b_{23} }{8} +\frac{b_{12} }{8} ,   
\end{equation} 
 and    \ben m_{3}^{\mathscr{E}} =m_{3} +\frac{b_{13} }{8} +\frac{b_{23} }{8} {~\rm for}~ 1\ne 2\ne 3. \en                                    
\item  For (\textit{aaa})\textit{-}type  baryons,
\begin{equation}    
m_{1}^{\mathscr{E}} =m_{2}^{\mathscr{E}} =  m_{3}^{\mathscr{E}} =m  +\frac{b_{12} }{4} ,                                              
\end{equation} 
and \ben  b_{12} =b_{23} =b_{13} {~\rm for} ~1=2=3.  \en
\end{enumerate}
 Values of quark masses and hyperfine interaction terms $b_{ij}$ are taken from (12) and (13), which in turn yield the following effective quark masses for $J^P =\frac{3}{2}^+$ baryons:
 \begin{enumerate}
\item  For singly heavy baryons,
                                   \[ m_{u}^{\Sigma _{_{b} }^{*} } =m_{d}^{\Sigma _{_{b} }^{*} } = 387 {~\rm MeV}, m_{b}^{\Sigma _{_{b} }^{*} } =5046 {~\rm MeV};\]
                                   
\[ m_{u}^{\Xi_{b}^{*} } = m_{d}^{\Xi _{b}^{*} } = 377 {~\rm MeV}, m_{s}^{\Xi_{b}^{*} } =555 {~\rm MeV}; m_{b}^{\Xi _{b}^{*} } =5044 {~\rm MeV};\]
\ben m_{s}^{\Omega _{b} ^{*}} =549 {~\rm MeV}, m_{b}^{\Omega _{b}^{*} } =5044 {~\rm MeV}.    \en
\item  For doubly heavy baryons,  
\[ m_{u}^{\Xi _{cb}^{*} } = 366 {~\rm MeV}, m_{c}^{\Xi _{cb}^{*} } =1714 {~\rm MeV}, m_{b}^{\Xi _{cb}^{*} } =5044 {~\rm MeV};\]
 
\[  m_{s}^{\Omega _{cb}^{*} } =542 {~\rm MeV}, m_{c}^{\Omega _{cb}^{*} } =1713 {~\rm MeV}, m_{b}^{\Omega _{cb}^{*} } =5044 {~\rm MeV};\]
  
  \[  m_{u}^{\Xi _{bb}^{*} } =m_{d}^{\Xi _{bb}^{*} } =364 {~\rm MeV}, m_{b}^{\Xi _{bb}^{*} } =5044 {~\rm MeV}; \]            
             \ben  m_{s}^{\Omega _{bb}^{*} } =540 {~\rm MeV}, m_{b}^{\Omega _{bb}^{*} } =5044 {~\rm MeV};  \en                                                  
\item  For triply heavy baryons,
\[ m_{c}^{\Omega _{_{ccb} }^{*} } =1711 {~\rm MeV}, m_{b}^{\Omega _{_{ccb} }^{*} } =5043 {~\rm MeV} \]

\[ m_{c}^{\Omega _{cbb}^{*} } =1710 {~\rm MeV}; m_{b}^{\Omega _{cbb} ^{*} } =5043 {~\rm MeV};   \]
\ben m_{b}^{\Omega _{bbb}^{*} } = 5043 {~\rm MeV}. \en
\end{enumerate}
 \begin{table}
\centering
\captionof{table} {Masses of $(J^P = \frac{3}{2}^+)$ bottom baryons using effective quark masses (in GeV).}
\label{mt2}
\begin{tabular} {|c |c |c|c|c|c|c|} \hline 
\textbf{Particles} & \textbf{Masses } &\textbf{\cite{30}} &\textbf{\cite{34}}&\textbf{\cite{38}}$^@$&\textbf{\cite{41}} &\textbf{Expt.\cite{2}}\\ \hline
\multicolumn{7}{|l|}{Singly heavy}   \\ \hline 
$\Sigma _{b}^{*} $ & 5.820 &5.882& 5.83$\pm$0.35 &5.878&5.834&5.833$^*$\\ \hline 
$\Xi _{b}^{*} $ & 5.976  & 5.975 &5.97$\pm$0.40&5.985&5.963 &5.9455 \\ \hline 
$\Omega _{b}^{*} $ & 6.142&6.063 & 6.08$\pm$0.40&6.116&6.088&- \\ \hline 

\multicolumn{7}{|l|}{Doubly heavy} \\ \hline
$\Xi _{cb}^{*} $ &  7.124 & 6.986 &7.25$\pm$0.20 & 6.921&6.980&- \\ \hline 
$\Omega _{cb}^{*} $ & 7.298 & 7.130 &7.30$\pm$0.20 & 6.997&7.130&- \\ \hline  
$\Xi _{bb}^{*} $ &  10.451 & 10.236 &10.40$\pm$0.10 & 10.219&10.237 &- \\ \hline 
$\Omega _{bb}^{*} $ & 10.628 &10.297 & 10.50$\pm$0.20 & 10.298&10.389 &- \\ \hline 
\multicolumn{7}{|l|}{Triply heavy} \\ \hline
$\Omega _{ccb}^{*} $ &  8.465 & - & -&8.181& -&- \\ \hline
$\Omega _{cbb}^{*} $ &  11.797 &- &-&11.488& -& -\\ \hline   
$\Omega _{bbb}^{*} $ &  15.129 &- &-&14.566&- & -\\ \hline 
\end{tabular}
\begin{tablenotes}
      \footnotesize 
      \item * used as input
      \item @ averaged
    \end{tablenotes}
\end{table}                                       
 We sum these effective quark masses to obtain masses of baryon iso-multiplets as shown in column 2 of Table \ref{mt2}. These masses are also compared with results of various approaches. We calculate the magnetic moments of $J^P=\frac{3}{2}^+$ baryons as given in column 2 of Table \ref{t4}. We compare our results with different works based on Bag model \cite{39}, NRQM \cite{12,39}  hyper central potential model \cite{38}, light cone QCD sum rules \cite{34}, and NRQM with AL1 potential model \cite{30}. The numerical results are discussed Section VI. 
 \begin{table}
\centering
\captionof{table} {Magnetic moments of $(J^P= \frac{3}{2}^+)$ bottom baryons using effective quark masses (in nuclear magneton).}
\label{t4}
\begin{tabular}{|c|>{\centering}p{2.6cm}|>{\centering}p{2.8cm}|c|c|c|c|c|} \hline 
\multirow{2}{*}{\textbf{ Baryons}} & \multicolumn{2}{|c|}{\textbf{This Work}} & \multirow{2}{*}{\textbf{ \cite{39} }} & \multirow{2}{*}{\textbf{ \cite{12, 39} }} & \multirow{2}{*}{\textbf{ \cite{38} }} & \multirow{2}{*}{\textbf{ \cite{34} }} & \multirow{2}{*}{\textbf{ \cite{30} }} \\ \cline{2-3} 
 &  \textbf{Effective Quark Mass}  & \textbf{ Screened Quark Charge} &  &  &  &  &  \\ \hline 
\multicolumn{8}{|l|}{Singly heavy} \\ \hline 
 $\Sigma _{b}^{*+} $  &  3.167  &  3.162 &  $2.346$  &  $3.56$  &  3.234  &  2.52$\pm $0.50  &  ---  \\ \hline 
 $\Sigma _{b}^{*0} $  &  0.746  & 0.705  &  $0.537$  &  $0.87$  &  0.791  &  0.50$\pm $0.15  &  ---  \\ \hline 
 $\Sigma _{b}^{*-} $  &  -1.677  &  -1.752 &  -$1.271$  &  -$1.92$  & -1.657  &  -1.50$\pm $0.36  &  ---  \\ \hline 
 $\Xi _{b}^{*0} $  &  1.031  &  0.915 &  $0.690$  &  $1.19$  &  0.042  &  0.50$\pm $0.15  &  ---  \\ \hline 
 $\Xi _{b}^{*-} $  &  -1.454  & -1.585  &  -$1.088$  &  -$1.60$  &  -1.098  & -1.42$\pm $0.35  &  ---  \\ \hline 
 $\Omega _{b}^{*-} $  &  -1.201  & -1.389 &-$0.919$     &  -$1.28$ &  -1.201  & -1.40$\pm $0.35   &  ---  \\ \hline 
 
\multicolumn{8}{|l|}{Doubly heavy} \\ \hline 
 $\Xi _{cb}^{*+} $  &  2.011 & 2.022  & $1.414$ &  $2.19$  &  2.052  & ---  &  $2.270$  \\ \hline 
 $\Xi _{cb}^{*0} $  &  -0.551 &  -0.508 &  -$0.257$  &  -$0.60$  &  -0.568  & ---  &  -$0.712$  \\ \hline 
 $\Omega _{cb}^{*0} $  &  -0.274  & -0.309  &  -$0.111$  &  -$0.28$  &  -0.317  & ---  &  -$0.261$  \\ \hline 
  $\Xi _{bb}^{*0} $  &  1.596  &  1.507 &  $0.916$  &  $1.74$  & 1.577  & ---  &  $1.870$  \\ \hline 
 $\Xi _{bb}^{*-} $  &  -0.984  & -1.029  &  -$0.652$  &  -$1.05$  & -0.952  & ---  &  -$1.110$  \\ \hline 
 $\Omega _{bb}^{*-} $  &  -0.703  &  -0.805 &  -$0.522$  &  -$0.73$  &  -0.711  & ---  &  -$0.662$  \\ \hline 
\multicolumn{8}{|l|}{Triply heavy} \\ \hline 
$\Omega _{ccb}^{*+} $  &  0.670  &  0.703 &  $0.659$  &  $0.72$  &  0.651 & ---  &  ---  \\ \hline 
 $\Omega _{cbb}^{*0} $  &  0.242 &  0.225 &  $0.225$  &  $0.27$  &  0.216 & ---  &  ---  \\ \hline 
  $\Omega _{bbb}^{*-} $  &  -0.186  &  -0.198 &  -$0.194$  &  -$0.18$  &  -0.195  & ---  &  -$0.180$  \\ \hline 
\end{tabular}
\end{table}
\section{MAGNETIC MOMENTS WITH EFFECTIVE MASS AND SHIELDED QUARK CHARGE:}

Similar to the variation of the quark mass resulting from its environment, the charge of a quark inside a baryon may also be affected. For example, when a quark inside a baryon is probed by a soft photon, its charge may be screened due to the presence of the neighboring quarks \cite{16}. This effect is in some sense similar to the shielding of the nuclear charge of the helium atom due to surrounding electron cloud. We take the effective charge to be linearly dependent on the charge of the shielding quarks. Thus effective charge of quark, \textit{a}, in the baryon $B(a, b, c)$ is taken as \cite{16}:
\begin{equation}    
e_{a} ^{B}   =  e_{a}   +  \alpha _{ab} e_{b}   +  \alpha _{ac} e_{c}     , 
\end{equation} 
where $e_{a} $ is the bare charge of quark \textit{a}. Taking $\alpha _{ab} =\alpha _{ba} $ and invoking the isospin symmetry, we obtain the following constraints :
\[\begin{array}{l} {\alpha _{uu}   =  \alpha _{ud}   =  \alpha _{dd}   =\beta   ,} \\ {\alpha _{us}   =  \alpha _{ds}   =\alpha   ,} \end{array}\] 
\[\alpha _{ss  }   =  \gamma ;\] 
in charm sector:
\[    
\alpha _{uc}   =  \alpha _{dc}   =  \beta '  , 
\] 
\[\alpha _{sc}   =  \delta   ,\] 
\[\alpha _{cc}   =  \gamma ';\] 
in bottom sector:
\[    
\alpha _{ub}   =  \alpha _{db}   =  \beta ''  , 
\] 
\[\alpha _{sb}   =  \delta '   ,\] \[\alpha _{cb}   =  \gamma '',\] \[\alpha_{bb}= \zeta;\]
Using the SU(3) we get, \ben \alpha   =  \beta   =\gamma ;\en 
 \begin{equation}  
\beta '    =  \delta , ~\rm{and}~ \beta ''    =  \delta' .
\end{equation} 
 We can further reduce these parameters to \begin{equation}    
\gamma   =  \gamma '  =  \delta {~\rm and~} \gamma '' = \delta' ;
\end{equation}
  \ben    
  \delta = \delta '= \zeta.
  \en
  using SU(4) and SU(5) flavor symmetry which are badly broken.
Redefining the magnetic moment operator,
\begin{equation}    
\pmb{\mu }  =\sum _{i} \frac{e_{i}^{B} }{2m_{i} ^{\mathscr{E}} } \pmb{\sigma _{i} } ,  
\end{equation} 
we determine the baryon magnetic moments using the \textit{p, n} and $\Lambda $ moments as input, and fix the quark masses for numerical calculations: $m_{u} =m_{d} =$370 MeV, $m_{s} =$494 MeV, and $\alpha =$0.033. Here we keep $m_{c} =$1680 MeV and $m_b=$ 5.043 GeV. The obtained numerical values are given in column 3 of Tables \ref{t2} and \ref{t4}, and are correspondingly compared with various approaches.
 
 \section{Numerical Results and Discussions}

In this paper, we have used effective quark mass and screened quark charge scheme to predict the  magnetic moments of all the $J^{P} =  \frac{1}{2}^{+} $ and $J^{P} =  \frac{3}{2}^{+} $  baryons up to \textit{b} = 3. We have used iso-multiplet masses \textit{N}, $\Delta $, $\Lambda $, $\Lambda _{c} $, $\Lambda _{b} $ etc. as input to obtain effective quark masses inside a baryon for both spin-$\frac{1}{2}$ and spin-$\frac{3}{2}$ baryons. Using these effective quark masses we then predict the magnetic moments of the bottom baryons. Later, we also include the effect of screened quark charge to calculate the magnetic moments. The summary of results is presented as follows:

\subsection{Magnetic Moments of $\pmb{J^P=\frac{1}{2}^+}$ Baryons} 

Presently, no experimental values of magnetic moments are available for heavy baryon (charm and bottom) sector. One may expect them to be measured experimentally in near future as many interesting experimental results has been put forward recently \cite{1,2,3,4,5,6,7}. Theoretically, bottom baryon magnetic moments have been calculated using various approaches listed in coulumns 4-9 of Table \ref{t2}. We wish to point out that in order to compare results of various models care must be take of the notation of primed and unprimed states of singly and doubly heavy baryons. We observe the following:
\begin{enumerate}

 \item Our results for bottom baryons with one heavy quark are consistent with the predictions of the hyper central model \cite{38} and relativistic three quark model \cite{36}. However, the results obtained in NRQM \cite{12} and quark model based on power-law potential \cite{17} are roughly $10\% - 15 \%$ larger than our predictions with few exceptions.
 \item Comparison with improved bag model\cite{39} reveals that numerical values calculated in  this approach are in general smaller than all other approaches. The reason being that for heavy baryons the bag radii and center-of-mass motion corrections are smaller which inturn decrease the magnetic moment values. Following NRQM \cite{12}, they have also considered the effects of mixing in $\Xi^0_b$ and $\Xi^{'0}_b$ states resulting in different values in these cases.     
 \item In light cone QCD sum rules \citep{32,33}, available predictions i.e. $\mu _{\Xi^{-}_{b} } $= - (0.08 $\pm$ 0.02) n.m., $\mu _{\Xi^{0}_{b} } $= - (0.045 $\pm$ 0.005) n.m. are consistent with our results, however, it predicts larger magnetic moment value for $\mu _{\Lambda^{0}_{b} } $= - (0.18 $\pm$ 0.05) n.m. Comparing our results with recent calculation based on hypercentral approach by \cite{42} we find that there results are on larger side even when compared to a similar approach \cite{38}. 
 \item In doubly heavy baryon sector, numerical values of magnetic moments of $ \Xi _{bb}^{0}  $, $ \Xi _{bb}^{-}  $, $ \Omega _{bb}^{-} $ and $\Lambda^0_b$ are consistent with all the other approaches except for bag model \cite{39} and relativistic three quark model \cite{36} predictions which are smaller than our results.
 \item Comparison of magnetic moments of doubly heavy $ \Xi _{cb}^{+}  $, $ \Xi _{cb}^{'+},  $ $ \Xi _{cb}^{0}  $, $ \Xi _{cb}^{'0} $, and $ \Omega _{cb}^{0} $, $ \Omega _{cb}^{'0} $ states in different approaches show disagreements. This may be attributed to choice of wave functions in different models. Our results are in nice agreement with to the NRQM with AL1 potential \cite{30} and hyper central approach \cite{38}, however, they are marginally higher than predictions of relativistic three quark model \cite{36}. Also, it has been argued that mixing induced by color-hyperfine splitting may affect the magnetic moment values which has been include in improved bag model \cite{39} predictions.  
 \item Our results involving singly heavy magnetic transition $B_{1/2}^{\prime} \to B_{1/2}$  moments are in good agreement with NRQM approach \cite{12}. In fact, analysis based on light cone QCD sum rules \cite{32} predicts a similar value for transition magnetic moment $\mu_{\Sigma_{b} \Lambda_{b}}= (1.6 \pm 0.4) $ n.m. However, for doubly heavy baryon state transition magnetic moments our result are small in comparison to other approaches.
 \item Considering the fact that magnetic moments of doubly and triply heavy baryons are virtually governed by magnetic moments of heavy quarks, all the approaches give almost similar values of magnetic moments of triply heavy ($\Omega^+_{ccb}$ and $\Omega^0_{cbb}$) baryons.    
 \end{enumerate}

\subsection{Magnetic Moments of $\pmb{J^P=\frac{3}{2}^+}$ Baryons} 
 
 Likewise spin-$\frac{1}{2}$ baryon sector, in the absence of any experimental information, several theoretical approaches has been used to estimate relatively simpler case of $J^P=\frac{3}{2}^+$ \textit{b}-baryon magnetic moments as listed in coulumns 4-8 of Table \ref{t4}. We observe the following:
\begin{enumerate}

 \item For the case of singly heavy bottom baryons, our results are consistent with the predictions of the hyper central model \cite{38}, though smaller than estimates given by NRQM \cite{12}. 
 \item As observed in $J^P=\frac{1}{2}^+$ case, numerical results obtained by improved bag model \cite{39} are smaller than all approaches.     
 \item The magnetic moments calculated in light cone QCD sum rules \cite{34} are in nice agreement with our predictions, except for $ \Xi^{*0}_{b} $ magnetic moment value which is smaller than our prediction.  
 \item In doubly heavy baryon sector, as expected, our results are consistent with hyper central potential model \cite{38} but larger then bag model \cite{39} predictions. The numerical values of $ \Xi _{cb}^{*+}  $, $ \Xi _{cb}^{*0}  $, and $ \Xi _{bb}^{*0}  $ magnetic moments in this work are smaller in comparison to the NRQM \cite{12} and NRQM with AL1 potential \cite{30} predictions, while rest of the predictions seems consistent with these approaches.
 \item Here also, magnetic moments of triply heavy baryons namely $\Omega^{*+}_{ccb}$, $\Omega^{0}_{cbb}$ and $\Omega^{*-}_{bbb}$ acquire roughly similar values in all theoretical works. 
  
 \end{enumerate}
 We hope these results will motivate experimental and theoretical analyses in this direction in near future.
\begin{acknowledgments}
One of the authors, RCV thanks C.S. Kim, Department of Physics, Yonsei University, Seoul (Korea) and H.Y. Cheng, Institute of Physics, Academia Sinica, Taipei (Taiwan) for their hospitality, where part of this work was done.  The work was supported by the National Research Foundation of Korea (NRF)
grant funded by Korea government of the Ministry of Education, Science and
Technology (MEST) (No. 2011-0017430) and (No. 2011-0020333).
\end{acknowledgments}

\end{document}